\documentclass[twocolumn,superscriptaddress,showpacs,preprintnumbers,amsmath,amssymb,aps,pre]{revtex4}
\usepackage{graphicx}
\usepackage{dcolumn}
\usepackage{bm}
\begin{document}
\title{Natural entropy fluctuations discriminate similar looking electric signals emitted from systems of different dynamics\footnote{Published in Physical Review E {\bf71}, 011110 (2005).}}
\author{P. A. Varotsos}
\email{pvaro@otenet.gr}
\affiliation{Solid State Section, Physics Department, University of Athens, Panepistimiopolis, Zografos 157 84, Athens, Greece}
\affiliation{Solid Earth Physics Institute, Physics Department, University of Athens, Panepistimiopolis, Zografos 157 84, Athens, Greece}
\author{N. V. Sarlis}
\affiliation{Solid State Section, Physics Department, University of Athens, Panepistimiopolis, Zografos 157 84, Athens, Greece}
\author{E. S. Skordas}
\affiliation{Solid Earth Physics Institute, Physics Department, University of Athens, Panepistimiopolis, Zografos 157 84, Athens, Greece}
\author{M. S. Lazaridou}
\affiliation{Solid State Section, Physics Department, University of Athens, Panepistimiopolis, Zografos 157 84, Athens, Greece}
\begin{abstract}
Complexity measures are introduced, that quantify the change of the natural 
entropy fluctuations at different length scales in time-series emitted from systems operating far from 
equilibrium. They identify impending sudden cardiac death (SD) by analyzing fifteen minutes 
electrocardiograms, and comparing to those of truly healthy humans (H). These measures seem to be 
complementary to the ones suggested recently [Phys. Rev. E {\bf 70}, 011106 (2004)]
 and altogether enable the classification of individuals into three 
categories: H, heart disease patients and  SD. All the SD individuals,
  who exhibit critical dynamics, result in a common 
behavior.
\end{abstract}
\maketitle

\section{Introduction}
The problem of distinguishing electric signals which, although look to be similar, they are
 emitted from systems of different dynamics, still attracts a strong interest. 
Two characteristic cases of major practical importance are: First, Seismic Electric
Signals (SES) activities, which are low frequency ($\leq$1 Hz) signals of dichotomous
 nature that have been found in Greece\cite{VAR86A,VAR86B,VAR03x} and Japan\cite{UYE02} to precede earthquakes, may look to be similar
 to ``artificial'' noises (AN), which are electrical disturbances emitted
from nearby man-made sources. It has been argued\cite{VAR86A,VAR02,VAR03x} that
 SES activities are emitted when the stress reaches a {\em critical} value in the EQ focal area.
 Second, sudden cardiac death (SD), which is the primary cause of mortality in
the industrialized world\cite{BRA03}, may occur even if the electrocardiogram
 (ECG) looks to be similar to that of truly healthy (H) humans. Sudden cardiac arrest may also be considered as a
dynamic phase transition (critical phenomenon)\cite{VAR03y,ref8}.

Both cases have been treated in Ref.\cite{ref8}, but here we only focus on the second one. 
The time-series will be analyzed in the natural time-domain.
The natural time $\chi$ is introduced\cite{VAR01,VAR02} by ascribing to the $m$-th
 pulse of an electric signal consisting of $N$ pulses, the value $\chi_m=m/N$
 and the analysis is made in terms of the couple $(\chi_m, Q_m)$, where $Q_m$
 denotes the duration of the $m$-th pulse. The entropy $S$ in the natural
time-domain\cite{VAR01,VAR03}  is defined as
 $S=\langle \chi \ln \chi \rangle - \langle \chi \rangle \ln \langle \chi \rangle$,
where $\langle \chi \ \ln \chi \rangle=\sum_{k=1}^N p_k \chi_k \ln \chi_k$ , $\langle \chi \rangle=\sum_{k=1}^N p_k \chi_k$
and $p_k=Q_k/\sum_{n=1}^N Q_n$. It is {\em dynamic} entropy depending on the {\em sequential}
order of pulses\cite{ref8}.
Here we calculate the value of $S$ for a number of consecutive pulses and study
how it varies within the recording (i.e., using a time-window of certain length $N_w$
 sliding, each time by one pulse, through the whole time-series). Thus, for a window of length $N_w$, 
when starting from the $m_0$-th pulse, we have
$
S(m_0,N_w)=\langle \chi \ln \chi \rangle_w - \langle \chi \rangle_w \ln \langle \chi \rangle_w,
$
where $\langle \chi \ln \chi \rangle_w=\sum_{k=1}^{N_w} p_{k,w} \chi_{k,w} \ln \chi_{k,w}$ ,
 $\langle \chi \rangle_w=\sum_{k=1}^{N_w} p_{k,w} \chi_{k,w}$
with $p_{k,w}=Q_{m_0-1+k}/\sum_{n=1}^{N_w} Q_{m_0-1+n}$ and $\chi_{k,w}=k/N_w$.
 This variation  is quantified by the standard deviation $\delta S$ ($= \delta S_{N_w}$) of 
$\{ S(m_0,N_w), m_0=1,2, \ldots N-N_w \}$.  The value of $\delta S$ may change to 
a different value $\delta S_{shuf}$ when repeating the same calculation but after {\em shuffling} 
the $Q_m$ randomly. In Ref.\cite{ref8} we showed that a distinction between 
SD and H can be achieved when calculating both $\delta S_{shuf}$ and $\delta S$
at the {\em same} (time-window) length $N_w$ and then studying their ratio 
$\delta S_{shuf}/ \delta S$(which is labeled by $\nu$).
 Here we show that a similar distinction may be alternatively achieved 
if we introduce {\em appropriate} 
measures that quantify the $\delta S$-variability upon {\em changing} the time-window length
and, interestingly,  their
values approach the value of the Markovian case in SD, who exhibit critical
dynamics. Furthermore, we show that the measures 
suggested in this paper exhibit a certain type of complementarity when compared to 
those discussed in \cite{ref8}.

 \begin{figure}
\includegraphics{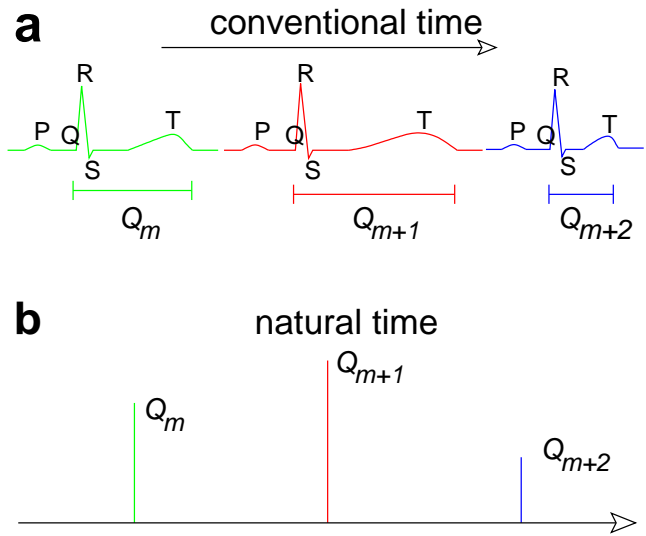}
\caption{\label{fig1} (Color) (a) Schematic diagram (not in scale)
of a three heartbeat excerpt of an ECG in the usual (conventional)
time-domain. Only the durations $Q_m$,$Q_{m+1}$,$Q_{m+2}$ of the
QT-interval (marked in each single cycle of the ECG corresponding
to one heartbeat) are shown. (b) The QT-interval time-series of
(a) read in natural time; the vertical bars are {\em equally}
spaced, but the length of each bar denotes the duration of the
corresponding QT-interval marked in (a).}
\end{figure}

In ECG, the turning points are traditionally labeled with the letters Q, R, S, T, 
see Fig.\ref{fig1}(a). (cf. In Fig. \ref{fig1}(b) we show, 
for example, how the QT interval time-series can be read in natural time).
 The RR- (beat-to-beat) and QRS-intervals (cf. mainly the RR)
can be automatically detected\cite{LAG90,LAG94,LAG97,JAN97} (which was followed here)
 more easily than the QT-. In spite of this fact, we intentionally study here all these 
three types of intervals for the following reasons: It has been clinically observed that the 
QT interval usually exhibits prolonged values before cardiac death (see Ref.\cite{KHA02} and references therein). 
Interestingly, this clinical observation was found\cite{ref8} to be consistent with the fact that
 in all SD the $\delta S$- (and $\delta S_{shuf}$-) values themselves of the QT-intervals
 exceed those of H, see Fig. \ref{fig2} (cf. the latter distinction between SD and H cannot be attributed
 to the allocation error of the QT interval, see Section VIII of Ref.\cite{epaps}).
Since the latter systematic behavior is not found when 
studying the RR- or the QRS- intervals\cite{ref8}, it is interesting to investigate
 here whether a systematicity occurs when employing the complexity measures suggested
 in this paper. Actually, we find that the latter measures seem to enable the
 distinction between SD and H when using the RR- and QRS- intervals of the original time-series.
 Furthermore, and most interestingly, we pinpoint that, even when solely using the
 most easily accessible values of the RR-intervals, such a distinction seems to
 be possible if we apply these measures to both the original time-series and the
 one obtained after shuffling the $Q_m$ randomly. We use here the QT Database
 from physiobank\cite{GOL00}, which includes fifteen minutes recordings of 
10 H and 24 SD (as well as recordings from four groups of heart disease patients, 
see below). Examples of the $\delta S$-values, calculated for the RR-, QRS- and QT-intervals
 in the range 3 to 100 beats are plotted in Figs. \ref{fig3}(a) and (b) for one H and one SD,
 respectively. As for the symbols, we use the same convention as in Ref.\cite{ref8}, i.e., 
$\delta S$ is used only when the calculation is made by a single time-window (e.g., 5 pulses),
 while the symbol $\overline{\delta S}$ stands for the average of the $\delta S$-values 
calculated for a sequence of single windows (e.g., 3, 4 pulses). 
Finally, $\langle \delta S \rangle$ denotes the $\delta S$-values averaged over a 
group of individuals, e.g., 10 healthy subjects.

Before proceeding, however, it might be useful to
 recapitulate the main differences of our procedure
 compared to several other earlier attempts by other groups.
 The reasons why the concept of entropy should be 
preferred (compared to other quantities) as 
discriminating statistics in physiological time-series have been explained in detail in Ref.\cite{ref8}.
 Furthermore, the advantages of using complexity measures based on 
{\em dynamic} entropy
 (and not on {\em static} entropy, e.g., Shannon entropy), as for example the Kolmogorov-Sinai 
entropy (K-S entropy), have been clarified\cite{ref8}. Earlier attempts in the ECG analysis 
have actually used measures related to dynamic entropy. For example, the so called 
approximate entropy (AE)\cite{PIN91} or sample entropy (SE)\cite{RIC00} 
have been introduced and later used by other authors (e.g., see Ref. \cite{NIK03} 
where AE is applied beyond other measures; see also Ref.\cite{epaps}). Also, 
Costa et al.\cite{COS02}
 introduced the multiscale entropy (MSE) approach, 
the algorithm of which is based on AE or SE, calculating the entropy
 at different scales. As for the $S$, which is also a dynamic entropy, 
as already mentioned, differs essentially from the other  ones, because it is defined\cite{VAR01,VAR03}
 in an entirely different time-domain (see Fig.\ref{fig1}(b)). Moreover, the following
 has been found: When studying the $S$-values themselves, most SES activities 
can be clearly distinguished\cite{VAR03} from the majority of AN, 
because they have S-values smaller and larger, respectively, than the value $S_u=0.0966$ 
of the ``uniform'' 
distribution (as the latter was defined in Refs.\cite{VAR03feb,VAR03}); 
on the other hand, when dealing with ECG they all have $S$-values comparable, more or less,
 to $S_u$ \cite{ref8}, see also \cite{epaps}, thus not allowing a 
clear distinction among their principal 
categories (cf. the entropy values themselves have been used in earlier attempts). 
This is achieved, however, when we quantify the $S$ {\em fluctuations}\cite{ref8} and use 
ratios of ``shuffled'' and ``unshuffled'' $S$ fluctuations on fixed time scales\cite{ref8} 
or ratios on different time scales that will be introduced here in Section II. Thus, in order
to discriminate similar looking electric signals emitted from systems of different dynamics,
the following seems to hold: signals that have $S$-values more or less comparable to $S_u$
(which is the case of all ECG) can be better classified by the
complexity measures relevant to the fluctuations $\delta S$ of the
entropy; if the $S$-values {\em markedly} differ from $S_u$ (which
is usually -but {\em not} always- the case of SES and AN), the
classification of these signals should be preferably made by the
use of the $S$-values themselves.

\section{THE NEW COMPLEXITY MEASURES PROPOSED. THE DISTINCTION BETWEEN SD AND H}

In classical Thermodynamics, the systems are studied close to equilibrium
and the relevant quantities have a clear physical meaning. In non-equilibrium systems,
 however, the meaning of entropy and its treatment should be handled with
great caution (e.g., \cite{VAR86A}), because there is at present
 (e.g., see Ref. \cite{BLY04}) no unified statistical mechanical theory underlying
 these systems. (cf. In transformations between non-equilibrium stationary states,
entropy might be a not well defined concept\cite{GAL03}; the connection of the entropy to
 microscopic dynamics is still a matter of intensive research (e.g.,\cite{TSA03} and references therein)).
In complex systems operating far from equilibrium (like the case
of heart dynamics\cite{GOL02}), long-range correlations
play an important role (cf. such correlations are, of course, of
prominent importance in equilibrium systems as well,
 when approaching a critical point, e.g., the ``critical''
temperature $T_c$, i.e., $T \rightarrow T_c$).
 Thus, in the latter
systems {\em both} correlations (i.e., short- and long-range), in general, is advisable to
be studied carefully and hence appropriate complexity measures should be envisaged. This
is, in simple words, the physics underlying the present paper and stimulated the procedure
 followed.

Along these lines, we introduce the ratios
$ \delta S_i(RR)/ \delta S_j(RR)$, $\delta S_i(QRS)/ \delta S_j(QRS)$ and
$\delta S_i(QT)/ \delta S_j(QT)$ for the RR-, QRS- and QT-intervals, respectively,
where $i,j$ denote the time-window length used in the calculation of $\delta S$. Assuming
 that $j<i$, these three ratios provide measures of the $\delta S$- variability when a
 scale $i$ changes to a scale $j$. We select as a common scale (for all RR-, QRS- and QT-)
the {\em smallest} $j$-value allowed for the natural time-domain analysis, i.e., $j=3$
 beats, and for the short range (s) $i=5$, while for the longer (L) $i=60$ beats.
 Thus, the following ratios are
 studied: $\lambda_s(\tau)\equiv \delta S_5(\tau )/ \delta S_3 (\tau )$ and $
\lambda_L(\tau)\equiv \delta S_{60} (\tau ) / \delta S_3(\tau )$,
where $\tau$ denotes the type of interval, i.e., $\tau=$RR, QRS or QT.
 We also  define the ratios  $\rho_i(\tau )=\delta S_i(RR)/ \delta S_i (\tau  )$,
which provide a {\em relative} measure of the $\delta S$-values of the RR-intervals
compared to either QRS- or QT- (for the {\em same} number of beats $i$). Here,
we will use for the short range   $\rho_s(\tau ) \equiv \rho_3(\tau )$ and
for the long range $\rho_L(\tau)\equiv \rho_{60}(\tau )$.

The calculated values for the complexity measures $\lambda_\kappa, \rho_\kappa$ (where $\kappa$ denotes
either the short, $\kappa=s$, or the longer, $\kappa=L$, range) are given, for all H and SD,
in Table \ref{tab1}. The minima $\min_{H}[\lambda_\kappa(\tau)]$ and maxima
$\max_{H}[\lambda_\kappa(\tau)]$ among the healthy individuals
for the RR ($\tau=$RR) and QRS ($\tau=$QRS) intervals
are also inserted in this Table.
We also include the corresponding minima $\min_{H}[\rho_\kappa(\tau)]$
and maxima $\max_{H}[\rho_\kappa(\tau)]$  for (the relative
$\delta S$-variability measure) $\rho$. For the sake of simplicity,
they  are labelled  $H_{min}$ and $H_{max}$, respectively
(and jointly named {\em H-limits}).
The superscripts `a' and `b' show the
cases of SD which have smaller and larger values than $H_{min}$ and $H_{max}$,
respectively.  In two individuals, i.e., sel41 and sel51, it is uncertain whether their
measure $\lambda_s(QRS)$ violates the value
 $H_{min}$=1.15.

\begin{figure}
\includegraphics{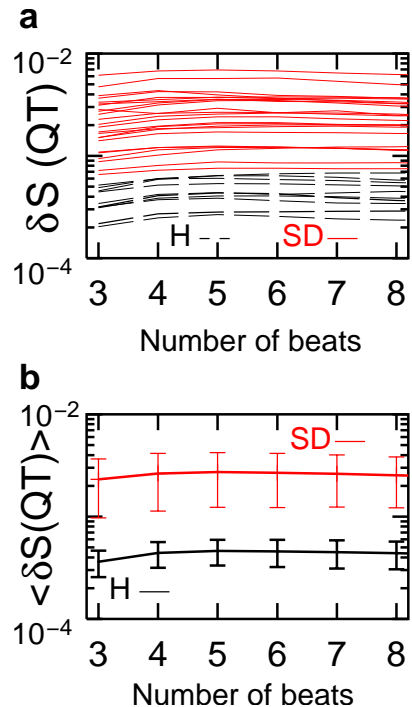}
\caption{\label{fig2} (color) (a) The $\delta S(QT)$-value for
each of the 24 SD and 10 H (see Table \ref{tab1}) and (b) the
average of the $\delta S$(QT) values -designated by $\langle
\delta S (QT) \rangle$- along with their standard error deviation
for each of the two groups SD and H versus the time-window
length.}
\end{figure}

\begin{figure}
\includegraphics{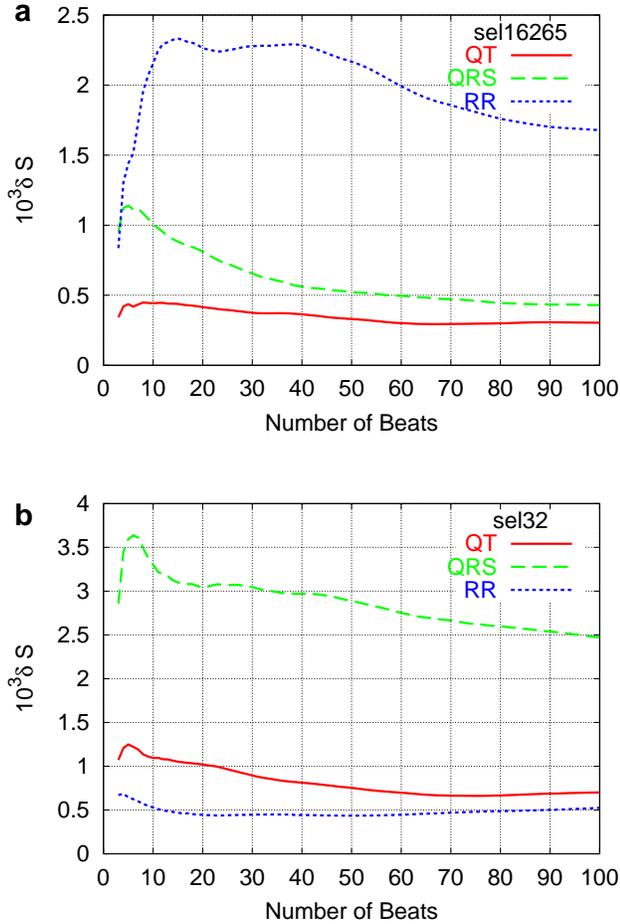}
\caption{\label{fig3} (Color) The $\delta S$-value versus the time-window length for one H
 (a) and one SD (b). Intervals: QT (solid red), QRS (broken green), and RR (dotted blue).}
\end{figure}

\begin{figure}
\includegraphics{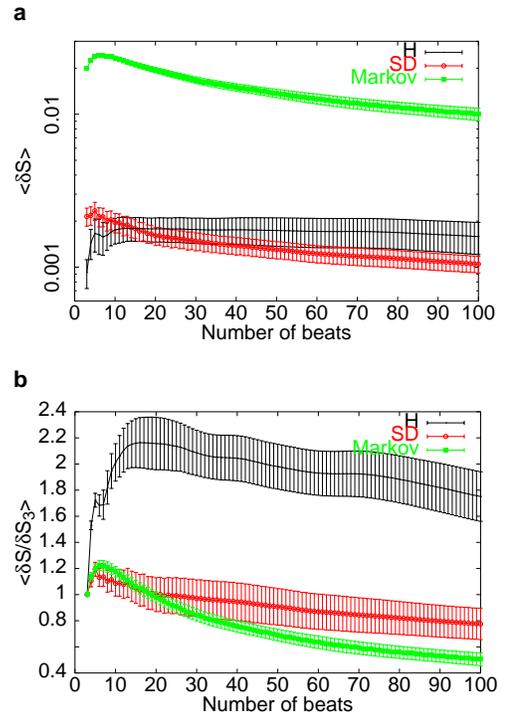}
\caption{\label{fig4} (Color) The average (denoted by the
brackets) values of (a): the
$\delta S (RR)$ and (b): $\delta S (RR)/ \delta S_{3} (RR)$  for the
SD (solid black) and H (red circles) versus the time-window length;
the bars correspond to the standard error of the mean.
The results for a Markovian time-series are also plotted (green squares),
but the bars here denote the standard deviation.  }
\end{figure}

\begin{figure}
\includegraphics{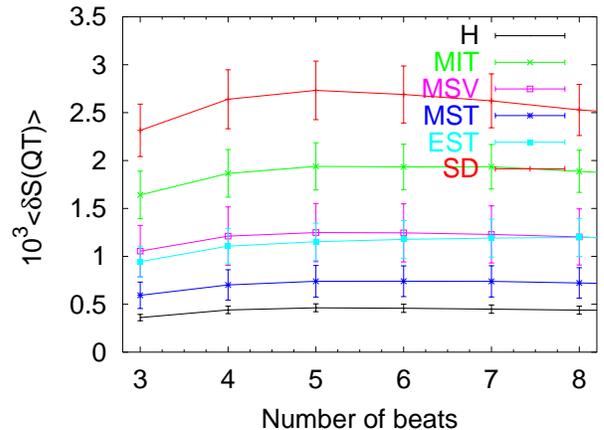}
\caption{\label{fig5} (color) The average of the $\delta S(QT)$
values -labelled $\langle \delta S (QT) \rangle$- for each of the
six groups labelled H, MIT, MSV, MST, EST and SD (see the text)
versus the time-window length. The bars denote the standard error
of the mean. (The corresponding standard deviations overlap
considerably and hence are not shown for the sake of clarity). The
lowermost- and the uppermost- curve correspond to H and SD,
respectively and hence coincide with the two curves depicted in
Fig. \ref{fig2}(b).}
\end{figure}

Table \ref{tab1} reveals that
 {\em all} SD violate one or more $H-$limits of $\lambda_s(RR)$, $\lambda_L(RR)$, $\rho_s(QRS)$ and $\rho_L(QRS)$,
and hence can be distinguished from H.
 In other words, the $\delta S$-variability measures of the RR-intervals, together
with their relative ones with respect to the QRS- (i.e., four parameters in total), seem to achieve a
distinction between SD and H. Note that  $\lambda_\kappa(RR)$ {\em
alone } can classify the vast majority of SD. Furthermore,
attention is drawn to the point that if we also consider the
$\lambda_\kappa (\tau )$ values calculated ( {\em not } in the
original, but) in the randomized (``shuffled'') sequence of $Q_m$,
we find that {\em all} SD violate one or more {\em H-limits} of $\lambda_\kappa ( RR )$ and
 $\lambda_{\kappa,shuf} ( RR )$ (see Table VII of Ref.\cite{epaps}). This  allows (cf. using again four
 parameters in total) the distinction of SD from H by using the RR-intervals {\em only}.

Thus, we found that among the 10 parameters defined in the original time-series extracted
from each ECG (or 20 parameters,
in total, if we also account for the corresponding parameters defined in the series obtained after
shuffling the $Q_m$ randomly), {\em only} four are required for the distinction between SD and H.
We clarify that this seems to be extremely difficult to be achieved by chance. In order to visualize it,
if we assume (for the sake of convenience only) independent and identically distributed (iid) distributions
of the parameters for one subject, we find that the probability that {\em all} 4 parameters are
within the bounds (minima and maxima) set by 10 other subjects (i.e., the healthy ones) is
$(1-2/11)^4\approx 0.448$.
Thus, the probability that all 24 additional subjects are classified as SD by pure chance is
$(1-0.448)^{24}\approx 6.4\times 10^{-7}$, i.e., extremely small. Concerning the validity 
of this statistical argument, we clarify that it  
does not remain valid if one just picks 4 parameters out of the original 20 ones. Only if one
decides which parameters wants to use {\em before} the calculation of the values, the argument is
 valid(This is the reason why blind evaluation- defining all methods, parameters and criteria studying 
one set of data, and {\em then} testing the significance using an additonal set of independent 
data- is considered very important in medical applications and/or publications).

We now attempt a physical interpretation of the present results,
the main feature of which focuses on the fact that both ratios
$\lambda_s(RR)$ and $\lambda_L(RR)$ become
smaller, in the vast majority of SD, compared to H. Recall that the $\delta S(RR)$-values
 themselves cannot
distinguish SD from H, see Fig. \ref{fig4}(a), in contrast to the
ratios $\delta S_i(RR) / \delta S_3 (RR)$, see Fig. \ref{fig4}(b).
Before proceeding, we mention two points: First, for
individuals at high risk of sudden death, fractal organization
(long range correlations) breaks down (see Refs.\cite{IVA01,GOL02}
and references therein). The breakdown of fractal physiologic
complexity is often accompanied by emergence of {\em uncorrelated
randomness} or {\em excessive} order (e.g., periodic oscillations
appear in the heart rate recordings of
``frequency''$\approx$1/min, which are associated with
Cheyne-Stokes breathing) \cite{GOL02}. Second, if we calculate\cite{ref8,VAR03} the
$\delta S$-values in a (dichotomous) Markovian (${\cal
M}$) time-series (exponentially distributed pulses), for a total
number of N=$10^3$ pulses (i.e., length comparable to that of the
ECG analyzed here), we find that these values: (a) lead to
$\lambda_s({\cal M})=1.20\pm0.03$ and (b) differ drastically, see Fig.
\ref{fig4}(a), from the $\delta S(RR)$-values themselves of {\em
both} SD and H (thus indicating that they exhibit non-Markovian
behavior on the whole; this is consistent with the
aspects that bodily rhythms, such as the heart beat, show complex
dynamics, e.g., \cite{IVA01,GOL02}). The fact that $\lambda_s(RR)$ in SD
becomes smaller than in H can now be understood as follows: Since
H exhibit a high order of complexity, it is expected that (even)
their $H_{min}$-value (=1.43), should markedly exceed
$\lambda_s({\cal M})$. On the other hand, in SD this high
complexity is lost, and hence their $\lambda_s(RR)$ values
naturally approach $\lambda_s({\cal M})$, thus becoming smaller.
Interestingly, the SD average value of $\lambda_s(RR)$ in Table
\ref{tab1} is 1.19, i.e., coincides with $\lambda_s({\cal M})$.
(Such a coincidence also occurs for the QRS-intervals in {\em
both} H and SD, which agrees with the observations\cite{KHA02} mentioned above
 that the prolonged
QT-intervals in SD mainly originate from enlarged ST-values,
 while their QRS- may remain the {\em same}.)
We now proceed to the interpretation of our results related to the
 ratio $\lambda_L(RR)$. In H, it is expected that
(in view of the RR long range correlations\cite{GOL02}) the
corresponding values must be appreciably larger than
$\lambda_L({\cal M})=0.64\pm0.05$, calculated in the Markovian
case (Fig. \ref{fig4}(b)). We now examine the SD: If, in SD ``{\em
uncorrelated randomness}'' appears, this reflects that their
$\lambda_L(RR)$ values naturally approach $\lambda_L({\cal M})$,
thus becoming smaller (compared to H); this actually occurs in the
vast majority of SD in Table \ref{tab1}. {\em If} in SD the
aforementioned  periodicities appear, it is naturally
expected to find {\em large} (see Ref.\cite{epaps}) $\delta
S$-values when a time-window of length around 60 beats, or so
(i.e., related to the aforementioned ``frequency''$\approx$1/min)
sweeps through the RR time-series, thus resulting in $\delta
S$-values even larger than those in H (since in H {\em no} such
periodicities appear). The latter  actually occurs in the few
cases marked with superscript `b' (i.e., those exceeding
$H_{max}$) in Table \ref{tab1} (For additional arguments on the
interpretation see \cite{epaps}).

The fact that the overall behavior of the complexity measures introduced
in this paper (i.e., clear distinction of SD from H) is more or less similar to that of the measures
 discussed in Ref.\cite{ref8}
 does not mean that the former measures are similar to the latter, because,
  as we shall explain below they exhibit a certain type of complementarity
  in the following sense: if in the frame of the one procedure an ambiguity emerges in the distinction between SD and H, the other procedure gives a clear answer.
  (Recall that, as mentioned in Section I, in Ref.\cite{ref8}
  we discussed entropy fluctuations -and ratios of ``shuffled'' and ``unshuffled''
   entropy fluctuations- on fixed time scales, while here we study
   entropy fluctuations on different time scales.) This is consistent with
the findings of Ashkenazy et al.\cite{ASH01} that an approach
dealing with ratios on the same time scale and an
approach dealing with ratios on different time scales
(or corresponding scaling exponents) are somewhat complementary.
 We now study, as an example, the following two procedures: i.e., the
 one that uses $\delta S$(QT)\cite{ref8} and the other which combines
 the measures $\lambda$,$\rho$.
 The -values of SD and H given in the last column of Table \ref{tab1} are
  classified into two classes: the larger values correspond to SD,
   and the lower ones correspond to H (see also Figs. \ref{fig2} and \ref{fig5} ).
    Let us focus on the two lowermost SD values and the uppermost H value.
     The former two correspond to sel33 and sel34 $\overline{\delta S}_{3-4}$(QT)=0.00076 and 0.00069, respectively) and the latter one to sel16795 ($\overline{\delta S}_{3-4}$(QT)=0.00056). In view of their
$\overline{\delta S}_{3-4}$(QT)-values proximity, one may wonder whether these two SD
could be confused with H. This ambiguity can be dissolved in the light of
the other procedure (i.e., $\lambda$, $\rho$), as follows: Table I reveals that
sel33 markedly violates both the $H_{min}$-limit for $\lambda_s$(QRS) as well as $H_{min}$ for $\lambda_s$(RR) (the latter can be visualized in Fig. \ref{fig6}). As for sel34, the $H_{max}$-limit of $\lambda_L$(QRS) is strongly violated. We now turn to an alternative example, i.e.,
 sel47, who, by means of the method using the complexity measures $\lambda$, $\rho$ (of the RR- and QRS-intervals), could be confused with H, because a deviation of only around 12\% from the $H_{min}$-limit of $\min_{H} [\rho_s(QRS)]=0.18$ is noticed. This ambiguity can be
 dissolved by means of the procedure using $\delta S$(QT) as follows: sel47 has $\overline{\delta S}_{3-4}$(QT)=0.0029 , which exceeds significantly, i.e., by a factor 5, the corresponding value
  of sel16795, who has the largest $\overline{\delta S}_{3-4}$(QT)=0.00056 value among the H.

\section{THE PROCEDURE TO DISTINGUISH SD FROM PATIENTS}

This Section aims at distinguishing SD from patients, where the latter terminology
 refers to individuals suffering only from heart diseases. The QT-Database of physiobank
  we use here, includes the following four groups of patients (a fifth group that consists
   of 4 individuals only, was disregarded for the reasons discussed in Ref.\cite{ref8}): 15 individuals from MIT-BIH Arrhythmia Database (labeled hereafter MIT), 13 from MIT-BIH Supraventricular
   Arrhythmia Database (MSV), 33 from the European ST-T Database (EST) and 6 from MIT-BIH
   ST change Database (MST). The values of $\lambda$, $\rho$, $\nu$,$\overline{\delta S}_{3-4}$(QT), $\lambda_{shuf}$, $\rho_{shuf}$ and $\overline{\delta S}_{3-4,shuf}$(QT) of all these patients are given in Ref.\cite{epaps}.

An inspection of the measures $\lambda$, $\rho$, $\nu$ shows three facts: First,
 all SD and all patients violate one or more $H$-limits. Second, {\em none} of the measures $\lambda$, $\rho$, $\nu$ alone, or a combination of two of them, can effectively differentiate the SD from the patients.
  Third, if we consider the three measures $\lambda$, $\rho$, $\nu$ (i.e., 16 parameters) altogether,
   we find that twenty SD out of 24  violate some of the limits of both patients and H, thus allowing in principle a distinction of the vast majority of SD from the other individuals.
   Thus, in summary, the consideration of the quantities ($\lambda$, $\rho$, $\nu$) only,
   does not lead to a distinction between {\em all} SD and patients. The same conclusion is drawn if we alternatively consider the quantities ($\lambda$, $\lambda_{shuf}$,$\rho$) only.

We now turn to the investigation of the $\delta S$(QT) values. In Fig. \ref{fig5}, the average
$\langle \delta S\rangle$(QT) value for each group is plotted versus the time-window length.
 It is intriguing that the results of the four groups (MIT, MSV, MST, EST) of patients
 are located between H (the lowermost curve) and SD (the uppermost curve). We emphasize,
  however, that if we plot the curves for each one of the 101 individuals
  (in a way similar to that of Fig.\ref{fig2}(a)), we find that there are
  some patients the results of which overlap with either SD or H. Let us consider
  only the limiting cases -i.e., the lowermost and the uppermost curve, to be called
  hereafter $\delta S$(QT)$_{min}$ and $\delta S$(QT)$_{max}$, respectively-
  obtained in each groups of patients. In order to distinguish SD from patients,
  we must appropriately discriminate the overlap which refers to those of the patients
   that lie above the uppermost $\delta S$(QT) curve of H; the latter curve from
    now on will be called $\delta S$(QT)$_{max,H}$. Thus, the limits
    of the patients we are currently interested in, do not
    extend from $\delta S$(QT)$_{min}$ to $\delta S$(QT)$_{max}$, since
    they must exceed $\delta S$(QT)$_{max,H}$, i.e.,
    \begin{equation}
   \delta S (QT) > \delta S(QT)_{max,H}.
   \label{con1}
   \end{equation}
The curve which corresponds to the one of the patients,
that has $\delta S$(QT) lying just above the $\delta S(QT)_{max,H}$ corresponds to a value, which will be labeled hereafter $\delta S$(QT)$_{min'}$. Thus, if we apply the condition
\begin{equation}
\delta S(QT)_{min'} \leq \delta S (QT) \leq \delta S(QT)_{max}
\label{con2}
\end{equation}
to each group of patients, we are left only with those of the patients that actually overlap with SD.

We now recall that, as mentioned above, the measures $\lambda$, $\rho$, $\nu$  altogether,
which are in fact ratios of $\delta S$ values, enable the discrimination
 of the vast majority of SD from all the others (i.e., patients and H),
  while the $\delta S$(QT) values themselves efficiently distinguish\cite{ref8} all SD from H. This motivates us to investigate whether a proper combination of these
   two facts can serve our purpose, which refers to the identification
   of all SD among the other individuals (patients and H). Thus, we now compare
    the quantities $\lambda$, $\rho$, $\nu$, $\delta S$(QT) altogether,
    of each SD, to the corresponding parameters of only
     those among the patients that happen to have $\delta S$(QT) values
exceeding the corresponding values of H, i.e., obey the condition (\ref{con1}), or preferably the more accurate condition (\ref{con2}). Such a comparison reveals that some
 of the 17 parameters of $\lambda$, $\rho$, $\nu$, $\delta S$(QT), in all SD,
 lie outside the limits of these patients (cf. the same happens, of course,
 if we compare each SD to the limits of H). These results point to the
 conclusion that all 24 SD are distinguished from the patients (and H).
 The same conclusion is drawn if we consider instead, the 17 parameters $\lambda$, $\lambda_{shuf}$, $\rho$, $\delta S$(QT). We emphasize, however, that the study of the
 estimation errors (see the Appendix) reveals that the confidence level for the distinction of
 all SD from the patients becomes appreciably larger if we combine all the measures $\lambda$, $\lambda_{shuf}$,  $\rho$ $\rho_{shuf}$, $\nu$ (of all intervals) with the condition (\ref{con2}) applied to both $\delta S$(QT) and $\delta S_{shuf}$(QT) (i.e., in reality, we then consider the
  limits of those patients whom {\em both} $\delta S$(QT)- and $\delta S_{shuf}$(QT)-values
  are larger than those in H, as shown in Fig. 6 of Ref.\cite{ref8}).

We finally comment on three points. First, once the identification of SD has been completed,
 the distinction between patients and H can be made by identifying as patients the individuals
 whom one or more of the aforementioned parameters violate the $H$-limits. Second, since it is known
 that heart rate variability depends strongly on the age, it is highly recommended that when
 comparing values of the aforementioned complexity measures, the corresponding limits should
 be taken from subjects (patients, H) of comparable age. Third, we now focus on the importance
 of the sequential order of $Q_m$ on the aforementioned complexity measures. We prefer to
 deal with the results related to the RR-intervals since it is known that the healthy heart
beats irregularly and that the intervals between beats (i.e., the RR-intervals) fluctuate
 widely, following complicated patterns\cite{CHI02}. Let us investigate, for example, the possibility
of using $\lambda_\kappa$(RR) alone to distinguish the SD as well as the four groups of
 patients from H, i.e., examine whether the $\lambda_\kappa$(RR)-values of each
 individual violate one (at least) of the relevant $H$-limits. The 
results show (see Table \ref{tab2}) that the vast majority of SD and
 of each group of patients is well distinguished from H 
by means of $\lambda_\kappa$(RR). The situation drastically changes,
 however, if we use, instead of $\lambda_\kappa$(RR), the $\lambda_{\kappa ,shuf}$-values
 (see the Tables V to VII in \cite{epaps}):
 only the minority of SD and of each group of patients can be differentiated
 from H. Since the calculation of the $\lambda_\kappa$(RR)-values takes into
 account the sequential order of $Q_m$, while the $\lambda_{\kappa ,shuf}$(RR)-values
 do not, this points to the following conclusion: It is the sequential order of beats
 that inherently contains the primary information which enables the distinction between
 the SD and patients, on the one hand, and the H, on the other. This might explain why
 procedures based on the entropy in natural time (which is dynamic entropy, affected by 
the sequential order\cite{VAR03y,ref8}, see Section I) -and hence they consider the complexity
 measures mentioned in the preceding Sections- can achieve such a distinction, while a static
 entropy (e.g., Shannon entropy, see Ref. \cite{ref8}) cannot.

\section{Conclusions}

First, in SD, the $\delta S$-values depend on the length scale in
a way significantly different from that in H. Hence these two groups
 of humans can be well distinguished. Second,  the SD, who exhibit
  critical dynamics, have $\lambda$-values (being, in fact,
 ratios of $\delta S$-values, as mentioned above) which approach those of the Markovian case.
  This should {\em not} be misinterpreted as showing that the corresponding time-series
   are of Markovian nature, because the $\delta S$-values themselves are
   approximately one order of magnitude smaller than those of
   the (dichotomous) Markovian time-series (see Fig. 4(a) and Ref. \cite{ref8}).
    Third, the quantities $\lambda$, $\lambda_{shuf}$, $\rho$, $\rho_{shuf}$, $\nu$, $\delta S$(QT) and
   $\delta S_{shuf}$(QT) {\em altogether}, seem to 
enable the classification of individuals into the three categories: H, patients and SD.

\appendix*
\section{THE INFLUENCE OF THE Estimation ERRORS ON THE PROCEDURES FOR THE DISTINCTION OF SD}
Beyond the error introduced by the use of an automatic threshold
detector for the allocation of the corresponding intervals (cf.
this is largest for the QT- and smallest for the RR-intervals),
the following two sources of errors must be considered\cite{VAR03y,ref8}: First, an
estimation error emerges when analyzing -instead of the original
time series of length $l \approx 10^3$- smaller lengths $l'$,
which, however, still significantly exceed the time-window lengths
used, for example $l' \approx 2 \times 10^2$ (the errors
associated with the measures in the short range,$s$, are smaller
from those corresponding to the longer range, $L$, because for the
latter range the $l/l'$ values -due to the restricted length of
the records available- are small, thus not allowing  more reliable
statistics). Second, a source of (statistical) error in the
results emerges when considering the ratio(s) $\delta
S_{shuf}$/$\delta S$ (i.e., when dealing with $\nu$ and
$\lambda_{shuf}$) instead of $\delta S$ itself. While $\delta S$
may be considered to have a {\em unique} value for a (given)
original $Q_m$ time-series, the value of $\delta S_{shuf}$ depends
on the randomly shuffled $Q_m$ series each time selected (cf. such
differences are well known\cite{kan02} when dealing with randomized series
of {\em finite} length). This is why the $\nu$-values given in
Ref.\cite{ref8} for SD and H do not fully coincide with those
tabulated in the present paper. To account roughly for the extent of this
statistical error, we averaged here the $\delta S_{shuf}$-values
calculated over a number (e.g. 20) of randomly shuffled
$Q_m$-series generated from the {\em same} original series and
then the corresponding standard deviation was estimated.

The final results on the above sources, could be summarized as
follows: The (percentage)
 estimation error was found to be around 10\% (cf. this is an {\em average} value)
 for the complexity measures $\lambda$,$\lambda_{shuf}$, $\rho$, $\rho_{shuf}$, $\nu$
 associated with the RR- and QRS-intervals.
Furthermore, since the error in the $\delta S$(QT) may reach 20\%,
the estimation error
 in those of the complexity measures that involve $\delta S$(QT) may be as high as $\approx$ 30\%.
Upon considering such error-levels, hereafter called ``plausible
estimation errors'' $\epsilon_p$, a study of each of the methods
for the distinction of SD was made. The study was repeated by
assuming larger (percentage) estimation errors, hereafter labeled
``modified estimation errors''$\epsilon_m$, calculated from
\begin{equation}
\epsilon_m=\epsilon_p \left( 1+\frac{H_{max}-H_{min}}{H_{max}+H_{min}} \right)
\end{equation}
for each parameter (see Table VIII in Ref. \cite{epaps}). Both
studies led,  more or less, to the same results. The
calculation, in each study, was made as follows: Each parameter
was assumed to be equal to its value (initially estimated from the
original time-series available) multiplied by a number randomly
selected in the range $1 \pm \epsilon_p$ or $1 \pm \epsilon_m$,
respectively) and then each of the methods for the distinction of
SD was applied. This application was repeated, for each method,
$10^3$ times via Monte Carlo and relevant conclusions have been
drawn for both studies. The extent to which these conclusions
hold, was also investigated in the following {\em extreme} case:
the limits of the parameters of H (and patients), which are
automatically adjusted for each ``random'' selection of the values
described above, have been assumed to {\em additionally} relax by
(extra) amounts equal to  $\epsilon_p$ or $\epsilon_m$. (Such a
``relaxation'' faces the {\em extreme} possibility that the
populations of H and patients treated here are not considered
large enough to allow a precise determination of their limits, and
hence future increased populations' studies could somehow broaden
these limits by  {\em extra} amounts as large as  $\epsilon_p$ or
$\epsilon_m$).

 The following conclusions were finally drawn concerning the 
distinction between SD and H (see also Table \ref{aca7}): Among the four
methods suggested (i.e., two in Ref.\cite{ref8} and two in Section II),
 the one that uses the
measures $\lambda$, $\rho$ (associated, however, with {\em all}
three types of intervals, i.e., 10 parameters in total) seems to
be robust in the following sense:  when assuming the error-levels
mentioned above, the use of $\lambda$, $\rho$ still allow with a
confidence level above 99\% the distinction of {\em all} SD from
H. ( Then a calculation similar to that given in Section II
concerning the probability that all 24 subjects are classified,
by means of 10 parameters, as SD
by {\em pure chance} -based on the limits set by 10 other subjects-
results in $[1-(1-2/11)^{10}]^{24}\approx 0.03$, i.e., too small.)
The confidence level decreases to 63\%, 49\%, 32\% and 59\% ,
respectively, when using four parameters or one parameter only as
follows: First: $\lambda_\kappa$(RR) and $\rho_\kappa$(QRS);
second: $\lambda_\kappa$(RR) and $\lambda_{\kappa,shuf}$(RR);
third: $\nu_\kappa$(RR) and $\nu_\kappa$(QRS); fourth:
$\overline{\delta S}_{3-4}$(QT). If we investigate the aforementioned  
extreme case of the additional ``relaxation'' of the {\em
H-limits}, the capability for
the distinction of {\em all} SD still remains with the following
results: In the case of using $\lambda$, $\rho$ (of all intervals),
the confidence level in distinguishing {\em all} SD is 88\%, while it
becomes {\em appreciably higher}, i.e., larger than 99\%, 
if we use the quantities $\lambda$,
$\rho$, $\lambda_{shuf}$, $\rho_{shuf}$, $\nu$, 
$\overline{\delta S}_{3-4}(QT)$, $\overline{\delta S}_{3-4, shuf}(QT)$ {\em altogether}. When
using, however, four parameters only in the first three
combinations mentioned above, the confidence level decreases to
90\%, 36\% and 8\% , respectively (and to 77\% when using
$\overline{\delta S}_{3-4}$(QT)), even when allowing two at the
most SD -out of 24- to be misinterpreted as being H.
As for the corresponding conclusions related to the distinction of SD from the patients,  
these can be drawn on the basis of the values given in the lower part of Table \ref{aca7}.
 
In summary, the study of the estimation errors reveals that
({\em if} the limits of the parameters will {\em not} be
broadened by future investigations) we can satisfactorily
distinguish the {\em totality} of SD from H as well as
discriminate the totality of SD from patients, 
upon employing the quantities  $\lambda$,
$\lambda_{shuf}$, $\rho$, $\rho_{shuf}$, $\nu$,$\overline{\delta
S}_{3-4}$(QT), $\overline{\delta S}_{3-4, shuf}$(QT) {\em
altogether},  i.e., the sixth and the last method, respectively in
Table \ref{aca7}. These quantities also allow the distinction of the 
{\em totality} of SD from H (as well as distinguishes the {\em vast majority}
of SD from the patients), {\em even if } their limits will be eventually 
broadened (by $\epsilon_m$).

The following remark should be  added concerning the number
of parameters required to achieve the desired distinction: In
reality, only twelve {\em independent} quantities, (i.e., the six:
$\delta S_{\kappa}(\tau)$ and the six $\delta
S_{\kappa,shuf}(\tau)$, where $\kappa$=s,L and $\tau$=RR,QRS, QT) 
are extracted from the
experimental data. Thus, for example, beyond $\overline{\delta S}_{3-4}(QT)$ 
{\em or} $\overline{\delta S}_{3-4,shuf}(QT)$, eleven 
additional parameters (out of 26) of the
ratios: $\lambda$, $\lambda_{shuf}$, $\rho$, $\rho_{shuf}$, $\nu$,
are in principle required to be used for the distinction. These
twelve quantities , however, should {\em not} be fortuitously
selected, but the following points must be carefully considered:
(i) priority should be given to the eight parameters associated
with $\lambda$-values and $\lambda_{shuf}$- (or $\nu$-) values of
RR and QRS, (ii) using, at least, one $\rho$-parameter (involving
$\overline{\delta S}_{3-4}(QT)$ or $\overline{\delta S}_{3-4,shuf}(QT)$), and (iii)
examining whether the totality of the parameters used can actually
reproduce the aforementioned twelve $\delta S$-values determined directly from
the data. However, in order to avoid the
difficulty arising from the completeness (or not) of the
aforementioned selection, at the present stage (i.e., until an
appreciably larger number of H and patients will be analyzed to
allow a better precision in the determination of the corresponding
limits), the preceding paragraph
 recommends to use -instead of  twelve- {\em all}   the
28 parameters associated with the quantities $\lambda$,
$\lambda_{shuf}$, $\rho$, $\rho_{shuf}$, $\nu$,
$\overline{\delta S}_{3-4}(QT)$ and $\overline{\delta S}_{3-4, shuf}(QT)$.

\bibliographystyle{apsrev}

\begin{turnpage}
\squeezetable
\begin{table*}
\caption{The variability measures ($\lambda$), the relative ones
($\rho$), and the ratios $\nu \equiv$ $\overline{\delta
S}_{shuf}/ \overline{{\delta S}}$ in the short (s) range and in
the longer (L) range in H (sel16265 to sel17453) and SD (sel30 to
sel17152) along with their $\overline{\delta S}_{3-4}(QT)$-values.} \label{tab1}
\begin{ruledtabular}

\begin{tabular}{c cc cc cc cc cc ccc ccc c}

  individual & RR &  & QRS & & QT & & RR over QRS & & RR over QT
\vspace{0.05cm} & & & 3-4 beats ($\nu_s$)\footnotemark[3] & & & 50-70 beats
($\nu_L$)\footnotemark[3] &
\\

 \cline{2-3}
 \cline{6-7}
 \cline{10-11}
\cline{15-17}

  & \vspace{0.05cm}$\lambda_s(RR)$ &
$\lambda_L(RR)$ &
 $\lambda_s(QRS)$
 & $\lambda_L(QRS)$ &
$\lambda_s(QT)$
 & $\lambda_L(QT)$ &
  $\rho_s(QRS)$ &
$\rho_L(QRS)$ & $\rho_s(QT)$ & $\rho_L(QT)$ &
 RR & QRS & QT & RR & QRS & QT & $\overline{\delta S}_{3-4}(QT) \times 10^{3}$\\
\hline
sel16265 &  1.72 &  2.38 &  1.19 &  0.52 &  1.27 &  0.88 &  0.88 &  4.01 &  2.44 &  6.62 &  1.87 &  0.98 &  1.29 &  0.48 &  1
.02 &  0.75 & 0.38\\
sel16272 &  1.69 &  1.35 &  1.29 &  0.61 &  1.21 &  0.50 &  0.18 &  0.40 &  0.67 &  1.79 &  1.65 &  0.88 &  0.94 &  0.77 &  1
.10 &  1.07 & 0.48\\
sel16273 &  1.61 &  2.69 &  1.16 &  0.59 &  1.30 &  1.11 &  1.11 &  5.05 &  3.17 &  7.65 &  2.18 &  0.99 &  1.46 &  0.50 &  0
.88 &  0.71 & 0.24\\
sel16420 &  1.51 &  1.74 &  1.22 &  0.48 &  1.37 &  0.66 &  0.96 &  3.46 &  1.97 &  5.21 &  1.60 &  0.99 &  1.07 &  0.53 &  1
.09 &  0.90 & 0.36\\
sel16483 &  1.43 &  2.37 &  1.23 &  0.49 &  1.31 &  0.68 &  0.25 &  1.22 &  0.96 &  3.37 &  2.27 &  0.99 &  1.17 &  0.52 &  1
.15 &  0.92 & 0.35\\
sel16539 &  2.00 &  1.94 &  1.26 &  0.50 &  1.41 &  1.08 &  1.85 &  7.10 &  5.57 & 10.04 &  1.43 &  1.07 &  1.27 &  0.50 &  1
.08 &  0.65 & 0.52\\
sel16773 &  1.92 &  2.61 &  1.21 &  0.49 &  1.31 &  0.70 &  0.90 &  4.84 &  1.49 &  5.54 &  1.85 &  1.01 &  0.91 &  0.44 &  1
.05 &  0.97 & 0.55\\
sel16786 &  1.71 &  1.57 &  1.19 &  0.51 &  1.31 &  0.84 &  1.16 &  3.56 &  3.97 &  7.43 &  1.39 &  1.01 &  1.19 &  0.55 &  1
.04 &  0.77 & 0.23\\
sel16795 &  1.77 &  0.99 &  1.24 &  0.55 &  1.16 &  0.56 &  0.77 &  1.37 &  2.87 &  5.08 &  1.10 &  0.98 &  1.05 &  0.74 &  0
.95 &  1.00 & 0.56\\
sel17453 &  1.87 &  1.67 &  1.26 &  0.54 &  1.22 &  0.68 &  1.49 &  4.59 &  2.91 &  7.12 &  1.46 &  1.01 &  1.02 &  0.57 &  0
.98 &  0.81 & 0.34\\
$H_{min}$ & 1.43 & 0.99 & 1.16 & 0.48 & 1.16 & 0.50 & 0.18 & 0.40 & 0.67 & 1.79 & 1.10 & 0.88 & 0.91 & 0.44 & 0.88 & 0.65 & 0.23\\
$H_{max}$ & 2.00 & 2.69 & 1.29 & 0.61 & 1.41 & 1.11 & 1.85 & 7.10 & 5.57 & 10.04 & 2.27 & 1.07 & 1.46 & 0.77 & 1.15 & 1.07 &
0.56\\
\\
sel30    &  1.11\footnotemark[1] &  0.89\footnotemark[1] &  1.20 &  1.05\footnotemark[2] &  1.28 &  0.56 &  0.51 &  0.43 &  1.73 &  2.73 &  1.15 &  1.08\footnotemark[2] &  1.13 &  0.66 &  0.71\footnotemark[1] &  1.10\footnotemark[2] & 1.04\footnotemark[2]\\
sel31    &  0.96\footnotemark[1] &  0.34\footnotemark[1] &  1.39\footnotemark[2] &  0.89\footnotemark[2] &  1.30 &  0.84 &  1.10 &  0.42 &  0.80 &  0.32\footnotemark[1] &  0.90\footnotemark[1] &  1.06 &  1.15 &  1.23\footnotemark[2] &  0.97 &  0.63\footnotemark[1] & 3.01\footnotemark[2]\\
sel32    &  0.96\footnotemark[1] &  0.67\footnotemark[1] &  1.26 &  0.96\footnotemark[2] &  1.16 &  0.65 &  0.23 &  0.16\footnotemark[1] &  0.63\footnotemark[1] &  0.64\footnotemark[1] &  1.31 &  1.11\footnotemark[2] &  1.13 &  1.02\footnotemark[2] &
  0.69\footnotemark[1] &  0.90 & 1.14\footnotemark[2]\\
sel33    &  1.14\footnotemark[1] &  0.77\footnotemark[1] &  0.96\footnotemark[1] &  0.52 &  1.21 &  0.53 &  0.79 &  1.17 &  2.41 &  3.50 &  1.07\footnotemark[1] &  1.00 &  1.08 &  0.85\footnotemark[2] &  0.83\footnotemark[1] &  1.00 & 0.76\footnotemark[2]\\
sel34    &  1.87 &  3.04\footnotemark[2] &  1.33\footnotemark[2] &  1.22\footnotemark[2] &  1.15\footnotemark[1] &  0.85 &  0
.40 &  1.00 &  1.16 &  4.12 &  2.13 &  1.11\footnotemark[2] &  1.12 &  0.41\footnotemark[1] &  0.77\footnotemark[1] &  0.67 &
 0.69\footnotemark[2]\\
sel35    &  1.12\footnotemark[1] &  0.52\footnotemark[1] &  1.24 &  0.66\footnotemark[2] &  1.12\footnotemark[1] &  0.44\footnotemark[1] &  1.72 &  1.36 &  0.83 &  0.99\footnotemark[1] &  1.02\footnotemark[1] &  0.97 &  0.97 &  1.02\footnotemark[2] &
  1.05 &  1.07 & 6.45\footnotemark[2]\\
sel36    &  1.31\footnotemark[1] &  0.62\footnotemark[1] &  1.12\footnotemark[1] &  0.51 &  1.26 &  0.60 &  2.35\footnotemark[2] &  2.88 &  1.45 &  1.52\footnotemark[1] &  1.03\footnotemark[1] &  1.01 &  1.08 &  0.93\footnotemark[2] &  0.99 &  0.89 &
 2.08\footnotemark[2]\\
sel37    &  0.92\footnotemark[1] &  0.71\footnotemark[1] &  1.26 &  0.87\footnotemark[2] &  1.11\footnotemark[1] &  0.78 &  0.71 &  0.58 &  1.19 &  1.07\footnotemark[1] &  1.11 &  1.17\footnotemark[2] &  1.07 &  0.56 &  0.75\footnotemark[1] &  0.64\footnotemark[1] & 3.30\footnotemark[2]\\
sel38    &  0.91\footnotemark[1] &  0.34\footnotemark[1] &  1.27 &  0.65\footnotemark[2] &  1.03\footnotemark[1] &  0.50 &  0.65 &  0.34\footnotemark[1] &  0.37\footnotemark[1] &  0.25\footnotemark[1] &  1.15 &  1.08 &  1.12 &  1.33\footnotemark[2] &
  0.89 &  1.03 & 2.71\footnotemark[2]\\
sel39    &  0.81\footnotemark[1] &  0.11\footnotemark[1] &  1.23 &  0.72\footnotemark[2] &  1.17 &  0.58 &  0.80 &  0.12\footnotemark[1] &  1.53 &  0.28\footnotemark[1] &  0.97\footnotemark[1] &  0.97 &  0.99 &  2.93\footnotemark[2] &  0.93 &  0.89 &
 2.44\footnotemark[2]\\
sel40    &  1.66 &  0.81\footnotemark[1] &  1.14\footnotemark[1] &  0.55 &  1.19 &  0.43\footnotemark[1] &  0.12\footnotemark[1] &  0.18\footnotemark[1] &  0.20\footnotemark[1] &  0.38\footnotemark[1] &  1.03\footnotemark[1] &  1.01 &  0.93 &  0.79\footnotemark[2] &  0.94 &  1.30\footnotemark[2] & 3.43\footnotemark[2]\\
sel41    &  1.14\footnotemark[1] &  0.48\footnotemark[1] &  1.18 &  0.70\footnotemark[2] &  1.22 &  0.56 &  0.21 &  0.15\footnotemark[1] &  0.80 &  0.68\footnotemark[1] &  0.91\footnotemark[1] &  1.04 &  1.06 &  1.05\footnotemark[2] &  0.84\footnotemark[1] &  0.96 & 1.53\footnotemark[2]\\
sel42    &  1.10\footnotemark[1] &  1.81 &  1.16 &  0.51 &  1.31 &  1.01 &  0.95 &  3.40 &  1.62 &  2.89 &  1.63 &  1.09\footnotemark[2] &  1.26 &  0.43\footnotemark[1] &  1.06 &  0.66 & 0.95\footnotemark[2]\\
sel43    &  1.69 &  3.04\footnotemark[2] &  1.24 &  0.77\footnotemark[2] &  1.26 &  0.68 &  0.06\footnotemark[1] &  0.23\footnotemark[1] &  0.11 &  0.48\footnotemark[1] &  2.79\footnotemark[2] &  1.12\footnotemark[2] &  1.08 &  0.56 &  0.77\footnotemark[1] &  0.89 & 2.23\footnotemark[2]\\
sel44    &  1.18\footnotemark[1] &  0.18\footnotemark[1] &  1.52\footnotemark[2] &  0.43\footnotemark[1] &  1.02\footnotemark[1] &  0.34\footnotemark[1] &  0.59 &  0.25\footnotemark[1] &  1.08 &  0.58\footnotemark[1] &  0.91\footnotemark[1] &  0.92 &
  0.90\footnotemark[1] &  2.25\footnotemark[2] &  1.46\footnotemark[2] &  1.33\footnotemark[2] & 4.12\footnotemark[2]\\
sel45    &  0.92\footnotemark[1] &  0.42\footnotemark[1] &  1.16 &  0.73\footnotemark[2] &  1.37 &  0.68 &  1.46 &  0.85 &  1.14 &  0.71\footnotemark[1] &  0.97\footnotemark[1] &  1.05 &  1.11 &  0.98\footnotemark[2] &  0.88 &  0.79 & 1.71\footnotemark[2]\\
sel46    &  0.94\footnotemark[1] &  0.43\footnotemark[1] &  1.05\footnotemark[1] &  0.71\footnotemark[2] &  1.12\footnotemark[1] &  0.55 &  1.35 &  0.82 &  1.59 &  1.26\footnotemark[1] &  1.01\footnotemark[1] &  0.99 &  1.01 &  0.99\footnotemark[2] &  0.85\footnotemark[1] &  1.01 & 3.44\footnotemark[2]\\
sel47    &  1.54 &  2.07 &  1.19 &  0.54 &  1.36 &  0.57 &  0.16\footnotemark[1] &  0.63 &  0.14\footnotemark[1] &  0.49\footnotemark[1] &  1.60 &  0.97 &  0.97 &  0.45 &  0.96 &  1.02 & 2.85\footnotemark[2]\\
sel48    &  0.84\footnotemark[1] &  0.30\footnotemark[1] &  1.23 &  1.08\footnotemark[2] &  1.14\footnotemark[1] &  1.00 &  0.91 &  0.26\footnotemark[1] &  1.36 &  0.41\footnotemark[1] &  0.84\footnotemark[1] &  1.24\footnotemark[2] &  1.42 &  1.49\footnotemark[2] &  0.68\footnotemark[1] &  0.74 & 1.75\footnotemark[2]\\
sel49    &  0.93\footnotemark[1] &  0.33\footnotemark[1] &  1.17 &  0.83\footnotemark[2] &  1.16 &  0.50 &  1.27 &  0.50 &  1.08 &  0.71\footnotemark[1] &  0.86\footnotemark[1] &  1.15\footnotemark[2] &  0.96 &  1.21\footnotemark[2] &  0.71\footnotemark[1] &  1.11\footnotemark[2] & 3.96\footnotemark[2]\\
sel50    &  1.32\footnotemark[1] &  0.59\footnotemark[1] &  1.28 &  0.46\footnotemark[1] &  1.21 &  0.32\footnotemark[1] &  1.78 &  2.31 &  1.21 &  2.26 &  1.07\footnotemark[1] &  1.00 &  0.91 &  0.93\footnotemark[2] &  1.20\footnotemark[2] &  1.62\footnotemark[2] & 5.21\footnotemark[2]\\
sel51    &  1.83 &  0.72\footnotemark[1] &  1.14\footnotemark[1] &  0.42\footnotemark[1] &  1.24 &  0.66 &  0.16\footnotemark[1] &  0.27\footnotemark[1] &  0.30\footnotemark[1] &  0.33\footnotemark[1] &  1.30 &  1.04 &  1.00 &  1.05\footnotemark[2] &
  1.24\footnotemark[2] &  0.90 & 1.83\footnotemark[2]\\
sel52    &  1.40\footnotemark[1] &  0.73 &  1.32\footnotemark[2] &  1.02\footnotemark[2] &  1.29 &  1.01 &  0.14\footnotemark[1] &  0.10\footnotemark[1] &  0.42\footnotemark[1] &  0.31\footnotemark[1] &  1.51 &  1.13\footnotemark[2] &  1.17 &  1.02\footnotemark[2] &  0.73\footnotemark[1] &  0.67 & 1.66\footnotemark[2]\\
sel17152 &  1.06\footnotemark[1] &  0.93\footnotemark[1] &  1.31\footnotemark[2] &  0.58 &  1.13\footnotemark[1] &  0.54 &  0.06\footnotemark[1] &  0.10\footnotemark[1] &  0.23\footnotemark[1] &  0.40\footnotemark[1] &  1.68 &  1.01 &  1.03 &  0.91\footnotemark[2] &  1.01 &  0.97 & 1.15\footnotemark[2]\\
$min$ & 0.81 & 0.11 & 0.96 & 0.42 & 1.02 & 0.32 & 0.06 & 0.10 & 0.11 & 0.25 & 0.84 & 0.92 & 0.90 & 0.41 & 0.68 & 0.63 & 0.69\\
$max$ & 1.87 & 3.04 & 1.52 & 1.22 & 1.37 & 1.01 & 2.35 & 3.40 & 2.41 & 4.12 & 2.79 & 1.24 & 1.42 & 2.93 & 1.46 & 1.62 & 6.45\
\

\end{tabular}
\footnotetext[1]{These values are smaller than the $H_{min}$ given
in each column}
 \footnotetext[2]{ These values are larger than the $H_{max}$
given in each column}
\footnotetext[3]{These values do not fully coincide with those given
in Ref.\cite{ref8} for the reasons discussed in the Appendix}
\end{ruledtabular}
\end{table*}
\end{turnpage}

\squeezetable
\begin{table}
\caption{The number of SD and patients that can be distinguished from H when
using $\lambda_\kappa$(RR) or $\lambda_{\kappa,shuf}$(RR) alone.} \label{tab2}
\begin{ruledtabular}
\begin{tabular}{c c c c c }

group & Total number &   $\lambda_\kappa$(RR) &   $\lambda_{\kappa,shuf}$(RR) &   $\lambda_\kappa$(RR) and $\lambda_{\kappa,shuf}$(RR) \\
\hline
SD & 24 & 23 & 10 & 24
\\
MIT & 15 & 14 & 6 & 14  \\
MSV & 13 & 13  & 2 & 13 \\
EST & 33 & 29 & 8 & 29  \\
MST & 6 & 5 & 0 & 5  \\
\end{tabular}
\end{ruledtabular}
\end{table}

\begin{turnpage}
\squeezetable
\begin{table*}
\caption{The confidence levels to distinguish SD from either H or
patients when considering the estimation errors $\epsilon_m$ discussed in the Appendix and given
in Table VIII of Ref.\cite{epaps}} \label{aca7}
\begin{ruledtabular}
\begin{tabular}{cccc ccc cccc}

  Method Employed & & & & Confidence levels to distinguish SD & &
  & & & & \\
 \cline{1-4}
 \cline{5-11}
 \\
 & & & & Using the limits &  & & & Using &
 &  \\
 & & & & from the data analyzed &  & & & broader limits\footnotemark[3] &
 &  \\
 \cline{5-7}
 \cline{8-11}
\\
Aim & Measures & Type of & No. of & All SD & All but & All but &
All SD & All but & All but & All but
\\
 &  & intervals & para- &  & one SD &
two SD\footnotemark[4] &  & one SD & two SD &
five SD\footnotemark[4] \\

 &  &  & meters & \% & \% &
\% & \% & \% & \% & \%
 \\
\hline Distinction & $\lambda$, $\rho$ & RR, QRS, QT & 10 & $>$99
& $>$99 & $>$99 &
 88 & 99 & $>$99 & $>$99 \\
of SD & $\lambda$, $\rho$ & RR, QRS & 4 & 63 & 95 & $>$99 & 8 & 43 & 90 & $>$99 \\
from H  & $\lambda$, $\lambda_{shuf}$ & RR & 4 & 49 & 90
& 99 & 1 & 11 & 36 & 97 \\
 & $\nu$ & RR,QRS & 4 & 32 & 74 & 96 & $<$0.5 & 1 & 8 & 60 \\
 & $\delta S_{3-4}(QT)$ & QT & 1 & 59 & 93 & $>$99 & 11 & 39 & 77 & $>$99 \\
& $\lambda$, $\rho$,$\lambda_{sh}$, $\rho_{sh}$, $\nu$, &  RR, QRS, QT & 28 & $>$99
& $>$99 & $>$99 &
 $>$99 & $>$ 99 & $>$99 & $>$99 \\
&$\delta S_{3-4}(QT)$,$\delta S_{sh, 3-4}(QT)$ & & & & & & & &  & \\
\\
Distinction & $\lambda$, $\rho$, $\nu$, $\delta
S_{3-4}(QT)$\footnotemark[1] & RR, QRS, QT & 17 & 51 &
 83 & 95 & $<$0.1 & $<$0.1 & $<$0.1 & 1 \\
of SD & $\lambda$, $\rho$, $\lambda_{sh}$, $\delta
S_{3-4}(QT)$\footnotemark[1] &
RR, QRS, QT & 17 & 62 & 91 & 98 & $<$0.1& $<$0.1&$<$0.1 & 1 \\
from patients  & $\lambda$, $\rho$, $\lambda_{sh}$, $\rho_{sh}$,
$\nu$, & RR, QRS, QT & 28 & 95 & $>$99 &$>$99 & 16 & 41
& 68 & 98 \\
 & $\delta S_{3-4}(QT)$, &  &  & &  & & & & &  \\
 & $\delta S_{sh, 3-4}(QT)$\footnotemark[2] &  &  & &  & & & & &  \\

\end{tabular}
\footnotetext[1]{Considering the limits of those patients that
have $\delta S_{3-4}(QT)$ larger than those in H}
\footnotetext[2]{Considering the limits of those patients that
have {\em both} $\delta S_{3-4}(QT)$ and $\delta S_{sh, 3-4}(QT)$
larger than those in H} \footnotetext[3]{by amounts $\epsilon_m$
given in Table VIII of Ref. \cite{epaps}} \footnotetext[4]{When stating, e.g., ``All but
one'', it means that when allowing {\em at the most}, one SD -out
of 24- to be misinterpreted as being H or patient, respectively}
\end{ruledtabular}
\end{table*}
\end{turnpage}

\begin{figure}
\includegraphics{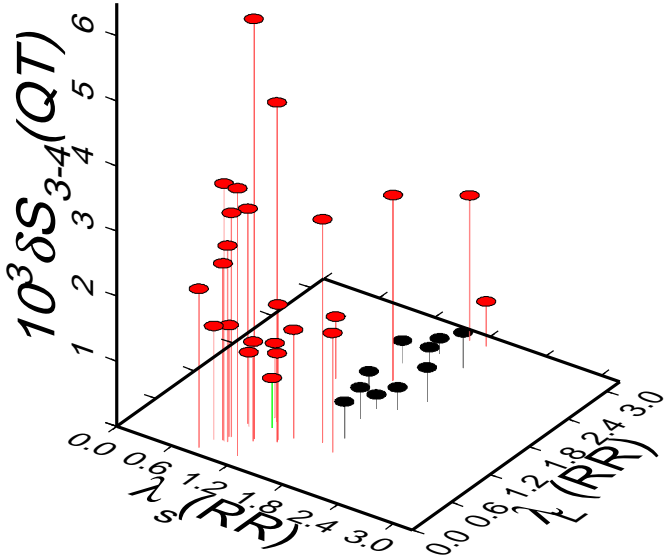}
\caption{\label{fig6} (color) The $\overline{\delta S}_{3-4}$(QT) values along with those of $\lambda_s$(RR)-
and $\lambda_L$(RR)- for SD (red) and H (black).
The individual sel33, who is discussed as an example in the text, is marked with a green column.   }
\end{figure}

\end{document}